\documentclass[prd,onecolumn,nofootinbib,showpacs,showkeys]{revtex4}

\usepackage{graphicx}
\def\bea{\begin{eqnarray}}
\def\eea{\end{eqnarray}}

\def\beq{\begin{equation}}
\def\eeq{\end{equation}}

\newbox\pippobox

\def\be{\begin{equation}}
\def\ee{\end{equation}}
\begin{document}
\input epsf
\title{Conformal Invariance in Einstein-Cartan-Weyl Space}
\author{Taeyoon Moon$^1$\footnote{dpproject@skku.edu},
Joohan
Lee$^2$\footnote{joohan@kerr.uos.ac.kr},
and Phillial
Oh$^1$\footnote{ploh@skku.edu}}
\affiliation{$^{1}$Department of Physics and Institute of Basic
Science, Sungkyunkwan University, Suwon 440-746 Korea}
\affiliation{$^2$Department of Physics, University of Seoul, Seoul
130-743 Korea}
\date{\today}
\begin{abstract}
 We consider  conformally invariant form of the actions in Einstein, Weyl, Einstein-Cartan and
 Einstein-Cartan-Weyl space in general dimensions($>2$) and investigate
 the relations among them. In Weyl space, the observational consistency condition for the vector field
 determining non-metricity of the connection can be obtained from the equation of motion.
 In Einstein-Cartan space
 a similar role is played by the vector part of the torsion tensor.
 We consider the case where the trace part of the torsion
 is the Kalb-Ramond type of field.
 In this case, we express conformally invariant action in terms of two scalar fields of
 conformal weight $-1$, which can be cast into some interesting form.
 We discuss some applications of the result.
\end{abstract}
,

\maketitle

\section{Introduction}

 The issues related to conformal transformation and conformal symmetry in Einstein's
general relativity have been studied for a long time. Einstein's
general relativity formulated in 1916 was successful in all known
experiments and it describes the reality very well. In spite of
these successes, there were some attempts to impose conformal
symmetry on Einstein's general relativity. One of the remarkable
attempts of these was made by Herman Weyl~\cite{weyl, weyl1, weyl2,
weyl3}. He proposed Einstein's metricity condition can be
generalized to incorporate
 conformal invariance ($\nabla_{\mu} g_{\alpha\beta}\sim
C_{\mu}g_{\alpha\beta}$, see next page for possible space-time
geometry).
 These attempts, however, had
a serious problem~\cite{violationclock, violationclock1,
violationclock2}. It is that quantum mechanics provides absolute
standards of length and Weyl's theory leads to observational
inconsistency, even though some authors suggested how to overcome
this inconsistency~\cite{clock}. Other
attempts~\cite{dirac,dirac1,cheng,hehl,pawl,nishino,demir,jackiw,jackiw1}
at incorporating conformal invariance in the theory of general
relativity have been studied in the context of particle physics and
mathematical physics. Recently, conformal invariant gravity was
suggested as the model of dark energy in~\cite{jain, jain1} and
conformal quintessence model was introduced
in~\cite{qu1,qu0,qu2,qu3}.


 On the other hand, one of the most famous theory for a generalizing Einstein's
 theory is
 the Einstein-Cartan theory~\cite{cartan, cartan1, cartan2}. In 1920s,
 $\acute{\rm{E}}$lie
 Cartan  suggested that space-time with torsion can be related to the intrinsic
 angular momentum,
 before the concept of spin
 was introduced. In the early 1960s, Sciama~\cite{sciama,sciama1} and
Kibble~\cite{kibble} reinterpreted Cartan's theory as the theory of
gravitation with spin and torsion. According to them, in order to
incorporate spinors into the theory of general relativity,
vierbein(or tetrad) must be introduced. This theory describes
General Relativity in terms of gauge theory with some local gauge
transformations, such as the local Poincare
group~\cite{sciama,sciama1,kibble,localpoincare, localpoincare1,
localpoincare2, localpoincare3, localpoincare4}.
Since then, torsion have been widely studied in general
relativity~\cite{shapiro,shapiro1,shapiro2,shapiro3}. Some author
showed that torsion tensor corresponds to Kalb-Ramond
field~\cite{kalb} in Einstein-Cartan space with Weyl's non-metricity
condition (hereafter referred to as Einstein-Cartan-Weyl
space)~\cite{scherk} and the traceless part of contorsion tensor to
Kalb-Ramond field in Einstein-Cartan space~\cite{sa}. In cosmology,
some of the applications include torsion quintessence,~\cite{capo},
possible role of torsion in current accelerating
universe~\cite{shie} and early inflation~\cite{boeh}.

In particular, the relation between torsion and conformal symmetry
was studied by several authors.  It was shown that the torsion could
play an important role in conformal invariance of the action and
behave like an effective gauge field in~\cite{neto, netoo}. Also, in
the non-minimally coupled metric-scalar-torsion theory, for some
special choice of the action, torsion acts as a compensating field
and the full theory can be conformally equivalent to general
relativity on classical level \cite{german,neto1}.

In this paper, we consider the local conformal invariance in the
Einstein-Cartan-Weyl space and explicitly construct an action with
local conformal invariance. We only pay attention to conformal
invariance at classical action level and we do not discuss a view
point of frame (Einstein frame or Jordan frame)\footnote{In a
conformally symmetric theory all choices of frames are equivalent.
The change of the conformal frame corresponds to the invariant
choice of the dynamical variables at the classical level. However,
when matters are coupled in such a way that the conformal invariance
is no longer valid, a choice of frame becomes very important. As
discussed in Ref. \cite{cho} a different choice of frame leads to a
different physical interpretation of the theory. In this work we do
not deal with such issues because we are only dealing with
conformally invariant theories. }~\cite{cho,gunzig}. In order to
achieve the conformal invariance, we introduce scalar fields which
are obtained through some special ansatz of the Weyl gauge field and
the trace part of the torsion. For Weyl gauge field, this ansatz is
natural because it can solve the problem of observational
inconsistency.\footnote{There also exist
literatures~\cite{smalley,dereli, dereli1} considering conformal
invariance in Einstein-Cartan-Weyl space. However, in these papers,
this problem was not considered. In our approach, the Weyl gauge
field is assumed to have the form of gradient scalar (see
eq.(\ref{c2})) from the beginning.} For the torsion vector field, it
is consistent with the equations of motion in Einstein-Cartan space.
The role of these scalar fields are
 some kind of gauge field which give conformal symmetry in the
 theory.
We first construct the conformal-invariant actions in Einstein,
Weyl, Einstein-Cartan space  in general dimensions and search for
relations among them. It can be shown
 that these actions are all equivalent to each other through the
 aforementioned ansatz.
Then, we extend the analysis to Einstein-Cartan-Weyl space and
explicitly construct conformally invariant action. Here, we end up
with two scalar fields coming from the Weyl's gauge field and
torsion vector field.  One of the motivation is that these two
fields could constitute the two scalar fields of the quintom model
\cite{quintom,quintom11,quintom12,quintom1} and provide a possible
geometric origin of the dark energy \cite{dark,dark1,dark2}. We will
examine the conformal symmetry
 in each of the four spaces with the following types of connection :\\
\\
A. metric compatible and torsionless $\bf{(Einstein~ space)}$: $
~\nabla_{\mu} g_{\alpha\beta}=0,~~\Gamma_{\mu\nu}^\rho=
\left\{_{\mu\nu}^\rho
\right\}.\nonumber$\\
B.  Weyl's type and torsionless $\bf{(Weyl ~space)}$:
$~\bar{\nabla}_{\mu} g_{\alpha\beta}\sim C_{\mu}
g_{\alpha\beta},~~\bar{\Gamma}_{\mu\nu}^\rho=\bar{\Gamma}_{(\mu\nu)}^\rho.\nonumber$\\
C.  metric compatible with torsion $\bf{(Einstein-Cartan~ space)}$:
$~\tilde{\nabla}_{\mu}
g_{\alpha\beta}=0,~~\tilde{\Gamma}_{\mu\nu}^\rho=
\left\{_{\mu\nu}^\rho
\right\}-K_{\mu\nu}^{\ \ \rho}.\nonumber$\\
D. Weyl's type with torsion $\bf{(Einstein-Cartan-Weyl~ space)}$:
$~\hat{\nabla}_{\mu} g_{\alpha\beta}\sim C_{\mu}
g_{\alpha\beta},~~\hat{\Gamma}_{\mu\nu}^\rho=\bar{\Gamma}_{(\mu\nu)}^\rho
-K_{\mu\nu}^{\ \ \rho}.\nonumber$\\

This paper is organized as follows. In Sec. II, we review Weyl's
theory and discuss the conditions for Weyl's vector field $C_{\mu}$
to avoid the observational consistency problem. In Sec. III, we
introduce Einstein-Hilbert action with conformal symmetry and
discuss and relations with Weyl action. In Sec. IV, Einstein-Cartan
action with conformal symmetry is considered and we will show that
this action link to other actions. In the course of constructing the
conformal-invariant action, we consider two cases. One is to take
the trace part of torsion as a physical field
 itself, the other is as an anti-symmetric tensor $B_{\mu\nu}$ of Kalb-Ramond type. In Sec V, we consider
Einstein-Cartan-Weyl action with conformal symmetry. We briefly
summarize the results and discuss them in Sec VI.
\\
\\
$\textbf{Our notations.}$\\
\\
 The metric signature is $-,+,+...+$. The Riemann, Ricci tensor
and curvature scalar are given by the Christoffel symbols
$\Gamma^{\rho}_{\mu\nu}$ and the inverse metric $g^{\mu\nu}$
\begin{eqnarray}
R^{\rho}_{~\sigma\mu\nu}&=&\partial_{\mu}\Gamma^{\rho}_{\nu\sigma}
-\partial_{\nu}\Gamma^{\rho}_{\mu\sigma}+
\Gamma^{\rho}_{\mu\lambda}\Gamma^{\lambda}_{\nu\sigma}-
\Gamma^{\rho}_{\nu\lambda}\Gamma^{\lambda}_{\mu\sigma},\nonumber\\
R_{\mu\nu}&=&\partial_{\rho}\Gamma^{\rho}_{\nu\mu}
-\partial_{\nu}\Gamma^{\rho}_{\rho\mu}+
\Gamma^{\rho}_{\rho\lambda}\Gamma^{\lambda}_{\nu\mu}-
\Gamma^{\rho}_{\nu\lambda}\Gamma^{\lambda}_{\rho\mu}\nonumber\\
&=&R^{\rho}_{~\mu\rho\nu},\nonumber\\
R&=&g^{\mu\nu}R_{\mu\nu}. \nonumber
\end{eqnarray}
 The affine connection for A$\sim$D is $\Gamma^{\rho}_{\mu\nu},
 ~\bar{\Gamma}^{\rho}_{\mu\nu},~\tilde{\Gamma}^{\rho}_{\mu\nu},
~\hat{\Gamma}^{\rho}_{\mu\nu}$ respectively, i.e.,
\begin{eqnarray}
\nabla_{\mu}V_{\nu}=\partial_{\mu}V_{\nu}-\Gamma^{\rho}_{\mu\nu}V_{\rho},
\nonumber\\
\bar{\nabla}_{\mu}V_{\nu}=\partial_{\mu}V_{\nu}-\bar{\Gamma}^{\rho}_{\mu\nu}V_{\rho},
\nonumber\\
\tilde{\nabla}_{\mu}V_{\nu}=\partial_{\mu}V_{\nu}-\tilde{\Gamma}^{\rho}_{\mu\nu}V_{\rho},
\nonumber\\
\hat{\nabla}_{\mu}V_{\nu}=\partial_{\mu}V_{\nu}-\hat{\Gamma}^{\rho}_{\mu\nu}V_{\rho}.
\nonumber
\end{eqnarray}

\section{A Review of the Weyl's Theory : Introduction and
Observational problem}

 In 1918, Herman Weyl suggested an intuitive notion of gauge invariance~\cite{weyl, weyl1, weyl2,
weyl3}
 according to natural generalization of metricity condition used in Einstein's general
 relativity. He assumed that Einstein's metricity condition can be replaced by
explicitly imposing conformal symmetry.
\begin{eqnarray}
\bar{\nabla}_{\mu}{g}_{\alpha\beta}=-2f C_\mu
g_{\alpha\beta}.\label{metric}
\end{eqnarray}
${\rm where}~\bar{\nabla} ~{\rm{is~}}the~covariant~ derivative~ in~
Weyl ~space, ~f~\rm {  is ~nonzero ~constant} ~and ~C_\mu \rm{~is~
some~ vector~field}$. This condition is invariant with respect to
conformal transformations
\begin{eqnarray}
g_{\alpha\beta}\longrightarrow e^{2\omega}g_{\alpha\beta},\label{g}\\
C_{\mu}\longrightarrow
C_{\mu}-\frac{1}{f}\partial_{\mu}\omega.\label{c}
\end{eqnarray}

Consequently, for a vector transported around a closed loop by
parallel displacement both the direction and length can change, but
the angle between two transported vectors must be conserved. And his
initial intention is to identify new vector field $C_{\mu}$ with the
electromagnetic four potential itself. From the above condition
(\ref{metric}), one can form conformal objects as follows
\begin{eqnarray}
\bar{\Gamma}_{\alpha\beta}^\lambda &=& \left\{_{\alpha\beta}^\lambda
\right\}-f (C^{\lambda} g_{\alpha\beta}-C_{\alpha}
\delta^{\lambda}_{\beta}-C_{\beta} \delta^{\lambda}_{\alpha}),
\\{\bar R}&=&g^{\mu\nu}{\bar R}^{\rho}_{\mu\rho\nu}=g^{\mu\nu}(
\partial_\rho \bar{\Gamma}_{\nu\mu}^\rho - \partial_\nu
\bar{\Gamma}_{\rho\mu}^\rho + \bar{\Gamma}_{\rho\sigma}^\rho
\bar{\Gamma}_{\nu\mu}^\sigma - \bar{\Gamma}_{\nu\sigma}^\rho
\bar{\Gamma}_{\rho\mu}^\sigma)\nonumber\\
&=&R + (2-2n)f{\nabla}_\mu C^\mu + (n-2)(1-n)f^2C_\mu
C^\mu\label{r0} .
\end{eqnarray}
It is interesting to note that there exists 2nd Ricci tensor ${\bar
R}^{\rho}_{\rho\mu\nu}$ (some author refer it as ``homothetic
curvature''~\cite{scherk,tonnelat}) because in Weyl space, Riemann
curvature tensor has no the antisymmetric property for first two
indices any more, i.e., ${\bar R}_{\mu\nu\rho\sigma}\neq-{\bar
R}_{\nu\mu\rho\sigma}$. Defining the homothetic curvature as
$\tilde{R}_{\mu\nu}$, it can be written as
\begin{eqnarray}
{\tilde R}_{\mu\nu} &=& g^{\rho\sigma}{\bar
R}_{\rho\sigma\mu\nu}=\partial_{\mu}\bar
{\Gamma}_{\nu\rho}^{\rho}-\partial_{\nu}\bar{
\Gamma}_{\mu\rho}^{\rho}\nonumber\\
&=&nf(\partial_{\mu} C_{\nu}-\partial_{\nu} C_{\mu})\label{r1}\\
&\equiv& nfF_{\mu\nu}\label{f1}.
\end{eqnarray}
 Since conformal weight of R is $-2$, we can construct the most
general conformal action by introducing some scalar field $\varphi$
(conformal weight is $-1$) in 4D as follows~\cite{padm},
\begin{eqnarray}
S=\int  \,d^4x
\sqrt{-g}\left\{\frac{\alpha_{1}}{12}\varphi^2\bar{R}-\frac{\alpha_{2}}{2}(D_{\mu}\varphi)
(D_{\nu}\varphi)g^{\mu\nu}-\frac{\alpha_{3}}{4}
F_{\mu\nu}F^{\mu\nu}-\frac{\lambda}{4!}\varphi^4\right\},\label{l0}
\end{eqnarray}
where$~D_{\mu}=\nabla_{\mu}-fC_{\mu}~\rm{
and}~\alpha_1,~\alpha_2,~\alpha_3, \rm{~are~constants}.$ We have
added a $\lambda\varphi^4$ term which is also conformally invariant
in 4D. We can easily check the conformal invariance of the action
with respect to (\ref{g}), (\ref{c}) and $\varphi\rightarrow
e^{-\omega}\varphi$. But, we should note that $F_{\mu\nu}$ term in
this action or $\tilde{R}_{\mu\nu}$ term in (\ref{r1}) pauses some
problems of observational inconsistencies \cite{clock} We will come
back to this shortly after. From (\ref{r0}), we can rewrite
(\ref{l0}) as follows,
\begin{eqnarray}
S=\int  \,d^4x  \sqrt{-g} \left\{\frac{\alpha_{1}}{12}\varphi^2
R-\frac{\alpha_{2}}{2}(\nabla_{\mu}\varphi)
(\nabla^{\mu}\varphi)-\frac{1}{2}f^2(\alpha_{1}+\alpha_{2})
\varphi^2C^2+(\alpha_{1}+\alpha_{2})\varphi
c_{\mu}\nabla^{\mu}\varphi-\frac{\alpha_{3}}{4}
F_{\mu\nu}F^{\mu\nu}-\frac{\lambda}{4!}\varphi^4\right\}.\label{l1}
\end{eqnarray}
 Using the freedom of conformal invariance and particular choices of
 arbitrary constants $\alpha_{1\sim3}$, we can set $\varphi=\sqrt{3/4\pi G}$,
 $\alpha_{1}=-\alpha_{2}=1$
 and$~\alpha_{3}=1$. Then the above action (\ref{l1}) becomes
\begin{eqnarray}
S=\int  \,d^4x  \sqrt{-g} \left\{\frac{1}{16\pi G}R-\frac{1}{4}
F_{\mu\nu}F^{\mu\nu}-\frac{\lambda}{4!}\left(\frac{3}{4\pi
G}\right)^2\right\}.\label{l2}
\end{eqnarray}
 This is just Einstein-Maxwell action with cosmological constant!

 But, in 1918, Einstein rejected Weyl's theory~\cite{violationclock, violationclock1,
violationclock2}.
 Einstein pointed out that according to Weyl's
theory, the reading of an atomic clock would depend  not only on
space-time geometry
but also on the unit length of the measurement.
Consequently, Weyl's theory would disagree with well-known
observations~\cite{disagree}. Now, consider the length in Weyl
space, $L=g_{\mu\nu}l^{\mu}l^{\nu}$  \cite{clock}. Then, the change
of the length under an infinitesimal parallel transport
$dx^{\sigma}$ is
\begin{eqnarray}
dL&=&\bar{\nabla}_{\sigma}g_{\mu\nu}dx^{\sigma}l^{\mu}l^{\nu}\\
&=&-2f L C_{\sigma}dx^{\sigma}.
\end{eqnarray}
We have used the result of (\ref{metric}) in the second step. But,
the above result causes observational problems. For example, set two
clocks at a given space-time point P. If these two clocks travel to
another point Q through different paths $C_1$ and $C_2$, these two
clocks are not synchronized according to gravitational effect in the
theory of general relativity. This is the ``First clock effect''. In
Weyl space, we have an additional synchronization loss (``Second
clock effect'') due to variation of the unit length of measurements
by different paths (conservation of the unit length implies
$\nabla_{\sigma} g_{\mu\nu}=0$). To avoid ``Second clock effect''
problem, we have to impose the coincidence of the unit length of
measurements for both observers at P without reference to any path.
This implies that
\begin{eqnarray}
\int_{C_1}dL&=&\int_{C_2}dL\\
\Leftrightarrow\oint dL&=&0=-2f L \oint C_{\mu} dx^{\mu}.
\end{eqnarray}
 Using the Stokes theorem we obtain the following condition,
\begin{eqnarray}
\nabla_{\mu}C_{\nu}-\nabla_{\nu}C_{\mu}=0\label{f2}.
\end{eqnarray}
 Consequently, to keep the observational consistency, $F_{\mu\nu}$
term should be vanishing ($F_{\mu\nu}$ of (\ref{f1}) and
(\ref{l0})$\sim$(\ref{l1})) and $C_\mu$ must be a pure gauge. So the
homothetic curvature in Weyl's original action cannot describe the
electromagnetic interaction.

 Now, the solution of (\ref{f2}) can be written as
\begin{eqnarray}
C_{\mu}=\frac{2}{(n-2)f}\frac{\partial_{\mu}\phi}{\phi}\label{c2},
\end{eqnarray}
by introducing a scalar field $\phi$ which transforms as
\begin{eqnarray}
\phi\rightarrow e^{\frac{2-n}{2}\omega}\phi.
\end{eqnarray}

\section{conformal einstein-hilbert action and Weyl's action}

 {\bf case A}. $ ~\nabla_{\mu}
g_{\alpha\beta}=0,~~\Gamma_{\mu\nu}^\rho= \left\{_{\mu\nu}^\rho
\right\}~\rm{with}~S_{CEH}=S_{conformal~
Einstein-Hilbert}$.\\
\\
 The Einstein-Hilbert action of general relativity has the form
\begin{eqnarray}
S_{EH}=\int d^nx \sqrt{-g} R.\label{EH}
\end{eqnarray}
 This action, of course, is not invariant under conformal
 transformation of the metric, i.e., $g_{\mu\nu}\rightarrow
 e^{2\omega}g_{\mu\nu}$, because the volume element and the curvature
 scalar transform
 as~\cite{wald}
 \begin{eqnarray}
 d^nx\sqrt{-g}&\longrightarrow& d^nx e^{n\omega}\sqrt{-g},\\
 R&\longrightarrow&
 e^{-2\omega}(R-2(n-1)\nabla_{\gamma}\nabla^{\gamma}\omega-(n-2)(n-1)
 \nabla_{\gamma}\omega\nabla^{\gamma}\omega),
 \end{eqnarray}
 where $n$ is the space-time dimensions.

 In order to impose conformal symmetry on the action (\ref{EH}), we need something to cancel
 overall weight $e^{(n-2)\omega}$.
 Now, one can introduce the most general conformal-invariant action as
 follows
\begin{eqnarray}
S_{CEH} = \int  \,d^nx  \sqrt{-g} \left\{\phi^2
R+\frac{4(n-1)}{n-2}\nabla_{\gamma}\phi\nabla^{\gamma}\phi-\lambda\phi^{\frac{2n}{n-2}}\right\},\label{s3}
\end{eqnarray}
which is invariant w.r.t $g_{\mu\nu}\rightarrow
e^{2\omega(x)}g_{\mu\nu},~\phi\rightarrow
e^{\frac{2-n}{2}\omega(x)}\phi$ and $\lambda$ is some constant. This
is a well known action which is written by many
 authors~\cite{dirac,dirac1},\cite{jain,jain1,qu1,qu0,qu2,qu3},\cite{kell,man,man1}.
 Some authors sometimes refer this to {\em scale~invariant~gravity} or {\em
conformal gravity}.\footnote{Do not confuse with conformal gravity
included $\int d^4x ~C^{\mu\nu\rho\sigma}C_{\mu\nu\rho\sigma}$ where
$C_{\mu\nu\rho\sigma}$ is the Weyl tensor  which has only fourth
order derivative terms. We do not consider this action in this
paper.} Through variation of (\ref{s3}), we can obtain the field
equations for $g_{\mu\nu},~\phi$ as
\begin{eqnarray}
\delta_{g^{\mu\nu}}S&=&
\phi^2(R_{\mu\nu}-\frac{1}{2}g_{\mu\nu}R)+\frac{2n}{n-2}\nabla_{\mu}\phi\nabla_{\nu}\phi-
2\phi\nabla_{\mu}\nabla_{\nu}\phi+2g_{\mu\nu}\phi\nabla_{\gamma}\nabla^{\gamma}\phi
-\frac{2g_{\mu\nu}}{n-2}\nabla_{\gamma}\phi\nabla^{\gamma}\phi+\frac{\lambda
g_{\mu\nu}}{2}
\phi^{\frac{2n}{n-2}} ,\label{eom1}\\
\delta_{\phi}~S&=&\phi
R-\frac{4(n-1)}{n-2}\nabla_{\gamma}\nabla^{\gamma}\phi-\frac{n}{n-2}\lambda\phi^{\frac{n+2}{n-2}}=0.
\label{eom3}
\end{eqnarray}

 Here, if we
contract  eq.(\ref{eom1}) with $g^{\mu\nu}$, then eq.(\ref{eom1}) is
equivalent to eq.(\ref{eom3}). Thus, we have only one independent
equation (\ref{eom1}). This can be viewed as a result of $\phi$
being a pure gauge. Nevertheless, $\phi$ has still gauge degree of
freedom given by $\phi\rightarrow e^{\frac{2-n}{2}\omega(x)}\phi$.
If we gauge fix $\phi$ and choose $\phi=\phi_0=\sqrt{1/16\pi G}$,
the action (\ref{s3}) reduces to the Einstein-Hilbert action with a
cosmological constant.\footnote{Note that $F(\phi, R)$ gravity
without conformal invariance in general is equivalent to a system
described by the Einstein-Hilbert action plus scalar fields via
conformal transformation \cite{maeda,maeda1}} In this sense, the
usual Einstein gravity can be thought of as a gauge fixed version of
the conformally invariant action.

{\bf case B}. $ ~\bar{\nabla}_{\mu}
g_{\alpha\beta}=-2fC_{\mu}g_{\alpha\beta},~~\bar{\Gamma}_{\alpha\beta}^\lambda
= \bar{\Gamma}_{(\alpha\beta)}^\lambda=\left\{_{\alpha\beta}^\lambda
\right\}-f (C^{\lambda} g_{\alpha\beta}-C_{\alpha}
\delta^{\lambda}_{\beta}-C_{\beta} \delta^{\lambda}_{\alpha}).$

As explained in Section II, in order to avoid observational
inconsistency we choose $F_{\mu\nu}=0$. Then, the most general
conformal invariant action can be written as
\begin{eqnarray}
S_{WEYL}&=&  \int  \,d^nx  \sqrt{-g} \left\{\phi^2\bar{R}-\alpha
g^{\mu\nu}(D_{\mu}\phi)(D_{\nu}\phi)-\lambda\phi^{\frac{2n}{n-2}}\right\},
\end{eqnarray}
by introducing the scalar $\phi$ with conformal weight $(2-n)/2$ as
before, where
\begin{eqnarray}
D_{\mu}\phi=\nabla_{\mu}\phi-\frac{(n-2)f}{2}C_{\mu}\phi.
\end{eqnarray}
If $C_\mu$ field is being treated as auxiliary, it can be eliminated
through equations of motion which yields exactly eq. (\ref{c2}).
Substituting back into the above equation, we obtain
\begin{eqnarray}
S_{WEYL}=\int  \,d^nx \sqrt{-g} \left\{\phi^2
R+\frac{4(n-1)}{n-2}\nabla_{\gamma}\phi\nabla^{\gamma}\phi-\lambda\phi^{\frac{2n}{n-2}}\right\}.\label{w1}
\end{eqnarray}
 This action, as anticipated, is exactly equivalent to the action (\ref{s3}) up to a total derivative.
 Consequently, starting from the condition (\ref{c2}) in Weyl space, one can obtain conformal gravity action.

\section{einstein-cartan with conformal symmetry}

{\bf case C.} $~\tilde{\nabla}_{\mu}
g_{\alpha\beta}=0,~~\tilde{\Gamma}_{\mu\nu}^\rho=
\left\{_{\mu\nu}^\rho \right\}-K_{\mu\nu}^{\ \ \rho}$.

 Let us first consider metric compatibility condition,
\begin{eqnarray}
\tilde{\nabla}_\mu
g_{\alpha\beta}=\partial_{\mu}g_{\alpha\beta}-\tilde{\Gamma}_{\mu\alpha}^\lambda
g_{\lambda\beta}-\tilde{\Gamma}_{\mu\beta}^\lambda
g_{\alpha\lambda}=0.\label{metricc}
\end{eqnarray}
 From the above condition (\ref{metricc}), one can easily find the connection
 as $\tilde{\Gamma}_{\alpha\beta}^\gamma =
 \left\{_{\alpha\beta}^\gamma \right\}- K_{\alpha\beta}^{\ \
\gamma},$ where $K_{\alpha\beta}^{\ \ \gamma}$ is the contorsion
tensor, which is given in terms of the torsion tensor by
\begin{equation}
{K_{\alpha\beta}}^{\gamma} = - {S_{\alpha\beta}}^{\gamma} +
S_{\beta~\alpha}^{~\gamma} - S^{\gamma}_{~\alpha\beta}
\label{contorsin}
\end{equation}
with the  torsion tensor
$S_{\mu\alpha}^{~~\lambda}=\tilde{\Gamma}_{[\mu\alpha]}^\lambda$.

 It is important to note that $K_{\alpha\beta\gamma}$ is anti-symmetric for last two indices
 and $S_{\alpha\beta\gamma}$ anti-symmetric for first two
 indices.
 Now, we can generally decompose the contorsion tensor (\ref{contorsin}) into
 traceless and traceful parts as follows \cite{sa}
\begin{equation}
{K_{\alpha\beta}}^{\gamma} = {\tilde{K}_{\alpha\beta}}^{~~~\gamma} -
\frac{2}{n-1}\left( \delta_{\alpha}^{\gamma} S_\beta  -
g_{\alpha\beta} S^\gamma \right), \label{decompos}
\end{equation}
where $\tilde{K}_{\alpha\beta}^{\ \ \alpha}=0$
 and $S_\beta$ is
the trace of the torsion tensor, $S_\beta = S^{\ \
\alpha}_{\alpha\beta}$. This decomposition, of course, means that we
decompose torsion tensor as
\begin{eqnarray}
S_{\alpha\beta}^{~~~\gamma} = \tilde{S}_{\alpha\beta}^{~~~\gamma} -
\frac{1}{n-1}\left( \delta_{\beta}^{\gamma} S_\alpha  -
\delta_{\alpha}^{\gamma} S_\beta \right), \label{decomposit1}
\end{eqnarray}
where $\tilde{S}_{\alpha\beta}^{\ \ \alpha}=0$. Making use of the
connection $\tilde{\Gamma}_{\alpha\beta}^\gamma$ with
(\ref{decompos}),
  we can write curvature scalar as follows
\begin{eqnarray}
\tilde{R} = R - 4{\nabla}_\mu S^\mu -\frac{4(n-2)}{n-1}S_\mu S^\mu -
\tilde{K}_{\nu\rho\alpha} \tilde{K}^{\alpha\nu\rho},\label{r2}
\end{eqnarray}
where $R$ is the Riemann curvature scalar calculated from the usual
Christoffel symbols $\Gamma_{\alpha\beta}^\gamma$.

 Here, using scalar field $\phi$, we can construct the following conformal-invariant
 action
\begin{eqnarray}
\int \,d^nx  \sqrt{-g} \phi^2\left\{
 R - 4{\nabla}_\mu S^\mu -\frac{4(n-2)}{n-1}S_\mu S^\mu -
\tilde{K}_{\nu\rho\alpha}
\tilde{K}^{\alpha\nu\rho}-\lambda\phi^{\frac{2n}{n-2}}\right\}\label{Rbb}
\end{eqnarray}
is invariant under the following conformal transformations :
\begin{eqnarray}
g_{\mu\nu}&\rightarrow&
 e^{2\omega}g_{\mu\nu},~~\phi\rightarrow
e^{\frac{2-n}{2}\omega}\phi,\\
S_{\mu}&\rightarrow&
S_{\mu}+\frac{1-n}{2}\nabla_{\mu}\omega,\label{smu}\\
\tilde{K}_{\mu\alpha}^{~~~\lambda}&\rightarrow&
\tilde{K}_{\mu\alpha}^{~~~\lambda}.\label{kk}
\end{eqnarray}
 Taking the variation with respect to $S_{\mu}$ and $\tilde{K}_{\nu\rho\alpha}$
 then we obtain the following
 equation \cite{neto1}
\begin{eqnarray}
S_{\nu}=
\frac{n-1}{n-2}\frac{\nabla_{\nu}\phi}{\phi},~~\tilde{K}_{\nu\rho\alpha}=0,\label{an}
\end{eqnarray}
 which is consistent with the transformation law. From the above equation, the action (\ref{Rbb}) can be
expressed simply as
\begin{eqnarray}
\int \,d^nx  \sqrt{-g} \left\{\phi^2
R+\frac{4(n-1)}{n-2}\nabla_{\gamma}\phi\nabla^{\gamma}\phi
-\lambda\phi^{\frac{2n}{n-2}}\right\}\label{scartan}
\end{eqnarray}
up to a surface term.  This action is also exactly same with the
action (\ref{s3}) and
 (\ref{w1}). As a result, the torsion vector $S_{\mu}$ (\ref{an}) is equivalent to
 $C_{\mu}$ (\ref{c2})
in Weyl space, when $f=2/(n-1)$.

More generally, in 4D if we extend to two scalar fields as follows
\begin{eqnarray}
S&=&\int  \,d^4x \sqrt{-g}(a_1\phi_1^2+a_2\phi_1\phi_2+a_3\phi_2^2)\tilde{R}\nonumber\\
&=&\int  \,d^4x \sqrt{-g}(a_1\phi_1^2+a_2\phi_1\phi_2+a_3\phi_2^2)
\left(R -4 {\nabla}_\mu S^\mu  -\frac{8}{3}S_\mu
S^\mu-\tilde{K}_{\nu\rho\alpha}
\tilde{K}^{\alpha\nu\rho}\right),\label{scartan1}
\end{eqnarray}
then can find $S_{\mu}$ through the variation of it as
\begin{eqnarray}
S_{\mu}=\frac{3}{4}\left(\frac{2a_1\phi_1\nabla_{\mu}\phi_1+2a_3\phi_2\nabla_{\mu}\phi_2+
a_2\phi_1\nabla_{\mu}\phi_2+a_2\phi_2\nabla_{\mu}\phi_1}{a_1\phi_1^2+a_2\phi_1\phi_2+a_3\phi_2^2}\right).
\label{Smu1}
\end{eqnarray}
And 
we
introduce the following form
\begin{eqnarray}
\tilde{K}_{\nu\rho\alpha}=F(\phi_1,\phi_2)H_{\nu\rho\alpha},\label{kt}
\end{eqnarray}
where
$F(\phi_1,\phi_2)=(c_1\phi_1^2+c_2\phi_1\phi_2+c_3\phi_2^2)^{-1},
H_{\nu\rho\alpha}=\nabla_{\nu}B_{\rho\alpha}+\nabla_{\rho}B_{\alpha\nu}+
\nabla_{\alpha}B_{\nu\rho}$ and $B_{\mu\nu}=-B_{\nu\mu}$. Varying
for the field $B_{\mu\nu}$ in the action (\ref{scartan1}) then the
equation suggests a solution with scalar field
$T$~\cite{Bachas,Bachas1} as
\begin{eqnarray}
H_{\nu\rho\alpha}=\frac{1}{W_1F^2}k\epsilon_{\nu\rho\alpha\beta}\nabla^{\beta}T,
\end{eqnarray}
where $k$ is a constant,
$W_1=a_1\phi_1^2+a_2\phi_1\phi_2+a_3\phi_2^2$,
$\epsilon_{\nu\rho\alpha\beta}$ is the Levi-Civita tensor and
\begin{eqnarray}
T=\left\{\frac{1}{2}\ln{(b_1\phi_1^2+b_2\phi_1\phi_2+b_3\phi_2^2)}-G\ln{\phi_1}-H\ln{\phi_2}\right\}\label{T}.
\end{eqnarray}
Here, $b_{1\sim 3}$ are constants, $G+H=1$. It is important to note
that $\tilde{K}_{\nu\rho}^{~~\sigma}$ (\ref{kt}) is invariant for
conformal transformation, i.e., $g_{\mu\nu}\rightarrow
e^{2\omega}g_{\mu\nu}$, $\phi_{1,2}\rightarrow
e^{-\omega}\phi_{1,2}$. For simplicity we consider only
$W_1=F^{-1}$. In this case, substituting (\ref{Smu1})$\sim$(\ref{T})
into (\ref{scartan1}) and rearranging terms then we obtain
\newpage
\begin{eqnarray}
\int \,d^4x
\sqrt{-g}&&\left\{(a_1\phi_1^2+a_2\phi_1\phi_2+a_3\phi_2^2)R-A_1(\nabla_{\mu}\phi_1)^2-A_2(\nabla_{\mu}\phi_2)^2
-A_3\nabla_{\gamma}\phi_1\nabla^{\gamma}\phi_2-A_4\frac{\phi_1}{\phi_2}\nabla_{\gamma}\phi_1\nabla^{\gamma}\phi_2
\nonumber\right.\\
&&\left.-A_5\frac{\phi_2}{\phi_1}\nabla_{\gamma}\phi_1\nabla^{\gamma}\phi_2-A_6\frac{\phi_1^2}{\phi_2^2}(\nabla_{\mu}\phi_2)^2
-
A_7\frac{\phi_2^2}{\phi_1^2}(\nabla_{\mu}\phi_1)^2-A_8\frac{\phi_2}{\phi_1}(\nabla_{\mu}\phi_1)^2
- A_9\frac{\phi_1}{\phi_2}(\nabla_{\mu}\phi_2)^2\nonumber\right.\\
&&\left.+12k^2(a_1\phi_1^2+a_2\phi_1\phi_2+a_3\phi_2^2)
\frac{(b_1\phi_1\nabla_{\mu}\phi_1+b_3\phi_2\nabla_{\mu}\phi_2+
\frac{1}{2}b_2\phi_1\nabla_{\mu}\phi_2+\frac{1}{2}b_2\phi_2\nabla_{\mu}\phi_1)}{(b_1\phi_1^2+b_2\phi_1\phi_2+b_3\phi_2^2)}
\left(G\frac{\nabla^{\mu}\phi_1}{\phi_1}+H\frac{\nabla^{\mu}\phi_2}{\phi_2}\right)\nonumber\right.\\
&&\left.+\frac{3}{2}\frac{(2a_1\phi_1\nabla_{\mu}\phi_1+2a_3\phi_2\nabla_{\mu}\phi_2+
a_2\phi_1\nabla_{\mu}\phi_2+a_2\phi_2\nabla_{\mu}\phi_1)^2}{a_1\phi_1^2+a_2\phi_1\phi_2+a_3\phi_2^2}\nonumber\right.\\
&&\left.-6k^2(a_1\phi_1^2+a_2\phi_1\phi_2+a_3\phi_2^2)\frac{(b_1\phi_1\nabla_{\mu}\phi_1+b_3\phi_2\nabla_{\mu}\phi_2+
\frac{1}{2}b_2\phi_1\nabla_{\mu}\phi_2+\frac{1}{2}b_2\phi_2\nabla_{\mu}\phi_1)^2}{(b_1\phi_1^2+b_2\phi_1\phi_2+b_3\phi_2^2)^2}
\right\} ,\label{R10}
\end{eqnarray}
where
\begin{eqnarray}
A_{1}&=&a_{1}(6k^{2}G^{2}),~~~~~
A_2=a_3(6k^2H^2),~~~~\hspace*{0.1em}
A_3=a_2(12k^2GH),\nonumber\\
A_4&=&a_1(12k^2GH),~~ A_5=a_3(12k^2GH),~~
A_6=a_1(6k^2H^2),\nonumber\\
A_7&=&a_3(6k^2G^2),~~~~~ A_8=a_2(6k^2G^2),~~~~~\hspace*{-0.1em}
A_9=a_2(6k^2H^2).\nonumber
\end{eqnarray}
In the case of $k=0$, the coefficients $A_{1\sim 9}$ are all zero.
Then, the action (\ref{R10}) becomes
\begin{eqnarray}
S=\int  \,d^4x
\sqrt{-g}\left\{(a_1\phi_1^2+a_2\phi_1\phi_2+a_3\phi_2^2)R+
\frac{3}{2}\frac{(2a_1\phi_1\nabla_{\mu}\phi_1+2a_3\phi_2\nabla_{\mu}\phi_2+
a_2\phi_1\nabla_{\mu}\phi_2+a_2\phi_2\nabla_{\mu}\phi_1)^2}{a_1\phi_1^2+a_2\phi_1\phi_2+a_3\phi_2^2}\right\}.
\label{s10}
\end{eqnarray}
When $a_1=a_2=0,~a_3=1$ or $a_3=a_2=0,~a_1=1$, the above action
(\ref{s10}) reduces to that of only one scalar field $\phi$ in
(\ref{w1}), (\ref{scartan}). In this case, $S_{\mu}$ (\ref{Smu1}) is
equivalent to the one in
 (\ref{an}).

 In particular, setting $a_1=a_3=b_1=b_3=0,~a_2=b_2=1,~ G=-1/2,~ H=3/2$ and $k^2=1/4$ then
 the action (\ref{R10}) can be expressed simply like
\begin{eqnarray}
S&=&\int  \,d^4x \sqrt{-g}(\phi_1\phi_2
R+6\nabla_{\mu}\phi_1\nabla^{\mu}\phi_2)\label{s11}\\
&=&\int  \,d^4x \sqrt{-g} \left\{(\Phi_1^2-\Phi_2^2)R
+6\nabla_{\mu}\Phi_1\nabla^{\mu}\Phi_1-6\nabla_{\mu}\Phi_2\nabla^{\mu}\Phi_2\right\},\label{cq}
\end{eqnarray}
where $\Phi_1=(\phi_1+\phi_2)/2$ and $\Phi_2=(\phi_1-\phi_2)/2$.
This action is invariant for $g_{\mu\nu}\rightarrow
e^{2\omega}g_{\mu\nu},~\Phi_{1,2}\rightarrow e^{-\omega}\Phi_{1,2}$.
 As we have shown in Section III.
 conformally invariant action i.e., (\ref{s3}), (\ref{w1}) and (\ref{scartan}) is reduced to
  Einstein-Hilbert action when $\phi=\phi_0=\sqrt{1/16\pi G}$.
  Similarly, it is pointed out that the above action (\ref{cq}) can be
  reduced to superquintessence or conformal quintessence~\cite{qu1,qu0,qu2,qu3} when $\Phi_1=\sqrt{1/16\pi
  G}$. In the case of $\Phi_2=0$, of course, it is
  exactly same with the action (\ref{s3}), (\ref{w1}) and
  (\ref{scartan}).

  More generally, in (\ref{s11}) if we add extra terms according
to conformal transformation rule of $\hat{R}$, i.e.,
$\hat{R}\rightarrow e^{-2\omega}\hat{R}$ then
\begin{eqnarray}
S&=&\int  \,d^4x \sqrt{-g}\left\{\phi_1\phi_2
R+6\nabla_{\mu}\phi_1\nabla^{\mu}\phi_2-V(\phi_1,\phi_2)\right\}\\
&=&\int  \,d^4x \sqrt{-g} \left\{(\Phi_1^2-\Phi_2^2)R
+6\nabla_{\mu}\Phi_1\nabla^{\mu}\Phi_1-6\nabla_{\mu}\Phi_2\nabla^{\mu}\Phi_2-V(\Phi_1,\Phi_2)\right\},\label{quintom}
\end{eqnarray}
where
\begin{eqnarray}
V(\phi_1,\phi_2)&=&\alpha_1\phi_1^4+\alpha_2\phi_2^4+\alpha_3\phi_1^2\phi_2^2+\alpha_4\phi_1^3\phi_2+\alpha_5
\phi_1\phi_2^3,\nonumber\\
V(\Phi_1,\Phi_2)&=&\beta_1\Phi_1^4+\beta_2\Phi_2^4+\beta_3\Phi_1^2\Phi_2^2+\beta_4\Phi_1^3\Phi_2+\beta_5
\Phi_1\Phi_2^3,\nonumber\\
\beta_1&=&\alpha_1+\alpha_2+\alpha_3+\alpha_4+\alpha_5,\nonumber\\
\beta_2&=&\alpha_1+\alpha_2+\alpha_3-\alpha_4-\alpha_5,\nonumber\\
\beta_3&=&6\alpha_1+6\alpha_2-2\alpha_3,\nonumber\\
\beta_4&=&4\alpha_1-4\alpha_2+2\alpha_4-2\alpha_5,\nonumber\\
\beta_5&=&4\alpha_1-4\alpha_2-2\alpha_4+2\alpha_5.\nonumber
\end{eqnarray}
In the case of $\Phi_1=\sqrt{1/16\pi
  G}$, the action (\ref{cq}) can be reduced to conformal quintessence model with
  a power-law potential for $\Phi_2$~\cite{qu2,qu3}. In particular, regarding
  the above action ($\ref{quintom}$) as matter in Einstein space one can construct the
  action of conformal quintom
  model as
\begin{eqnarray}
S=\int  \,d^4x \sqrt{-g} \left\{R+(\Phi_1^2-\Phi_2^2)R
+6\nabla_{\mu}\Phi_1\nabla^{\mu}\Phi_1-6\nabla_{\mu}\Phi_2\nabla^{\mu}\Phi_2-V(\Phi_1,\Phi_2)\right\}.
\end{eqnarray}

\section{einstein-cartan-weyl theory with conformal symmetry}

{\bf case D.} $\hat{\nabla}_{\mu} g_{\alpha\beta}\sim C_{\mu}
g_{\alpha\beta},~~\hat{\Gamma}_{\mu\nu}^\rho=\bar{\Gamma}_{(\mu\nu)}^\rho
-K_{\mu\nu}^{\ \ \rho}$.

 Let us consider the extended conformal condition
\begin{eqnarray}
\hat{\nabla}_\mu g_{\alpha\beta}&=&-2fC_{\mu}g_{\alpha\beta}\nonumber\\
&=&\partial_{\mu}g_{\alpha\beta}-\hat{\Gamma}_{\mu\alpha}^\lambda
g_{\lambda\beta}-\hat{\Gamma}_{\mu\beta}^\lambda
g_{\alpha\lambda}.\label{hatg}
\end{eqnarray}
From the above condition (\ref{hatg}) one can find the connection as
follows
\begin{eqnarray}
\hat{\Gamma}_{\alpha\beta}^\gamma = \left\{_{\alpha\beta}^\gamma
\right\}-f(C^{\gamma}
g_{\alpha\beta}-C_{\alpha}\delta^{\gamma}_{\beta}-
C_{\beta}\delta^{\gamma}_{\alpha}) - K_{\alpha\beta}^{\ \ \gamma}.
\end{eqnarray}
 And the
curvature scalar calculated by using the connection
$\hat{\Gamma}_{\alpha\beta}^\mu$ is
\begin{equation}
\hat{R} = R + (2-2n)f{\nabla}_\mu C^\mu + (n-2)(1-n)f^2C_\mu C^\mu
-4\nabla_{\mu}S^{\mu}+4f(2-n)C_{\mu}S^{\mu}-\frac{4(n-2)}{n-1}S_{\mu}
S^{\mu}-\tilde{K}_{\nu\rho\alpha} \tilde{K}^{\alpha\nu\rho}.
\label{R4}
\end{equation}

 The above curvature scalar (\ref{R4}) is invariant with respect
 to~\cite{smalley}
\begin{eqnarray}
g_{\mu\nu}&\rightarrow& e^{2\omega}g_{\mu\nu},~~C_{\rho}\rightarrow
C_{\rho}-\frac{1}{f}\left(1-\frac{2\xi}{1-n}\right)\nabla_{\rho}\omega,\nonumber\\
S_{\mu}&\rightarrow&
S_{\mu}+\xi\nabla_{\mu}\omega,~~\tilde{K}_{\mu\alpha}^{~~~\lambda}\rightarrow
\tilde{K}_{\mu\alpha}^{~~~\lambda}\label{ECWT},
\end{eqnarray}
where $\xi$ is a constant. It is interesting to note that (\ref{R4})
can be written as the following simple form
\begin{eqnarray}
R - 4{\nabla}_\mu Y^\mu - \frac{4(n-2)}{n-1}Y_\mu Y^\mu
-\tilde{K}_{\nu\rho\alpha} \tilde{K}^{\alpha\nu\rho},
\end{eqnarray}
where $Y_{\mu}=(n-1)f C_{\mu}/2+S_{\mu}$. For
$Y_{\mu}\leftrightarrow S_{\mu}$, the above curvature scalar is
exactly same with the curvature scalar (\ref{r2}) in Einstein-Cartan
space and the conformal transformation rules
 for $S_{\mu}$ (\ref{smu}), $Y_{\mu}$ (\ref{ECWT}) are the same too, i.e.,
\begin{eqnarray}
Y_{\mu}\rightarrow Y_{\mu}+\frac{1-n}{2}\nabla_{\mu}\omega.
\end{eqnarray}
As a result, in Einstein-Cartan-Weyl space the action of curvature
scalar (\ref{R4}) can be reduced to (\ref{s3}), (\ref{w1}) and
(\ref{scartan}). In fact, the connection in Einstein-Cartan-Weyl
space can be written as $\hat{\Gamma}_{\alpha\beta}^\gamma =
 \left\{_{\alpha\beta}^\gamma \right\}- K_{\alpha\beta}^{\ \
\gamma}+2C_{\alpha}\delta_{\beta}^{\gamma}/(n-1),$
where $K_{\alpha\beta}^{\ \
\gamma}=
{\tilde{K}_{\alpha\beta}}^{~~~\gamma}
- 2\left(
\delta_{\alpha}^{\gamma}
Y_\beta  - g_{\alpha\beta}
Y^\gamma \right)/(n-1)$, and
the curvature scalar by this
connection is equivalent to the
curvature scalar (\ref{r2}) for
 $S_{\mu}\leftrightarrow Y_{\mu}$. This is because there is no contribution of
$2C_{\alpha}\delta_{\beta}^{\gamma}/(n-1)$
term in the course of
calculating curvature scalar.
In Eq. (\ref{ECWT}), proper
combination of $S_{\mu}$ and
$C_{\mu}$ is independent of
$\xi$ and the full action only
depends on this combination.

 The
fact that in the action only
$Y_{\mu}$ appears implies that
there is a new symmetry of the
form : $C_{\mu}\rightarrow
C_{\mu}+\alpha_{\mu}$ and
$S_{\mu}\rightarrow
S_{\mu}-(n-1)f\alpha_{\mu}/2$,
where $\alpha_{\mu}$ is
arbitrary function of the
space-time. The exact nature of
the symmetry requires further
investigation. Note that this
symmetry disappears when this
theory couples to external
matter because $C_{\mu}$ and
$S_{\mu}$ play different
geometric roles when
interacting with matter.

\section{conclusion and discussion}

 In the present paper, we studied conformally invariant actions in Einstein, Weyl,
 Einstein-Cartan and Einstein-Cartan-Weyl space and showed that these actions
 all have the same form. In particular, in Einstein-Cartan space it is shown that
 we can obtain conformally invariant action with two scalar fields. This is possible only
 for gradient scalar fields ansatz.
 It is natural that we take this ansatz in Weyl space because it solves the problem of observational
 inconsistency.
The torsion vector field is auxiliary and can be eliminated by the
equation of motion resulting in gradient scalar field.

We found that the conformally invariant action in Einstein-Cartan
space can be reduced to
 conformal quintessence model with a power-law potential. Also, one can construct the
 action whose matter part can be thought of as a conformal-quintom type.
 Quintom model has two scalar fields, and we pointed out that
 their geometric origin could come from the traceless part  and  trace part of
 the torsion tensor in Einstein-Cartan space. The conformal quintom model needs further
 investigation.

As was pointed out at the end of previous section, the degeneracy
between the torsion and Weyl gauge field in the Einstein-Cartan-Weyl
space can be lifted when external matter fields are introduced. It
would be interesting to elaborate on this aspect further and to
study the geometrical role of these fields in the presence of matter
fields.

In the conformally invariant formulation of the Einstein's action,
the scalar fields $\phi$ play the role of the conformal gauge field,
which is dynamically trivial. Therefore, such a formulation would be
interesting only if exact conformal invariance is broken by some
mechanism. Some of the possibilities is by quantum mechanical
effect, or by introducing some potential, or by using it as
conformal matter. However, it is important to remark that both of
the fields $\phi_{1,2}$ in the two scalar field case can not be
gauged away. It means that if we gauge away or fix one scalar field
from the theory, it becomes a gravity theory with a non-minimal
coupling of a scalar field. It would be interesting to check the
type of non-minimally coupled theory coming from the conformally
invariant action in detail.

\acknowledgements

 This work
was supported by the National
Research Foundation of
Korea(NRF) grant funded by the
Korea government(MEST) through
the Center for Quantum
Spacetime(CQUeST) of Sogang
University with grant number
2005-0049409.


\end{document}